\documentclass[useAMS,usenatbib]{mn2e}
\usepackage{graphicx}
\usepackage{enumerate}
\usepackage{epsfig}

\title[Pulse-phase resolved spectroscopy of SAX J1808.4-3658]{Pulse-phase resolved spectroscopy of continuum and reflection in SAX J1808.4-3658}
\author[T. Wilkinson, A. Patruno, A. Watts \& P. Uttley]{Tony Wilkinson$^{1}$\thanks{E-mail:
t.r.wilkinson@phys.soton.ac.uk}, Alessandro Patruno$^{2}$, Anna Watts$^{2}$ and Philip Uttley$^{1}$\\
$^{1}$School of Physics and Astronomy, University of Southampton, Southampton SO17 1BJ\\
$^{2}$Astronomical Institute ``Anton Pannekoek'', University of Amsterdam, Science Park 904, 1098 XH Amsterdam, Netherlands}
\begin{document}

\date{}

\pagerange{\pageref{firstpage}--\pageref{lastpage}} \pubyear{2009}

\maketitle

\label{firstpage}

\begin{abstract}

We perform phase-resolved spectroscopy of the accreting millisecond pulsar, SAX J1808.4-3658, during the slow-decay phase of the 2002 outburst. Simple phenomenological fits to RXTE PCA data reveal a pulsation in the iron line at the spin frequency of the neutron star. However, fitting more complex spectral models reveals a degeneracy between iron-line pulsations and changes in the underlying hotspot blackbody temperature with phase. By comparing with the variations in reflection continuum, which are much weaker than the iron line variations, we infer that the iron-line is not pulsed.  The observed spectral variations can be explained by variations in blackbody temperature associated with rotational Doppler shifts at the neutron star surface.  By allowing blackbody temperature to vary in this way, we also find a larger phase-shift between the pulsations in the Comptonised and blackbody components than has been seen in previous work. The phase-shift between the pulsation in the blackbody temperature and normalisation is consistent with a simple model where the Doppler shift is maximised at the limb of the neutron star, $\sim90^{\circ}$ prior to maximisation of the hot-spot projected area. 

\end{abstract}

\begin{keywords}
stars: neutron - X-rays: binaries - X-rays: individual (SAX J1808.4-3658) accretion
\end{keywords}

\section{Introduction}

SAX J1808.4-3658 was the first accreting millisecond X-ray pulsar (AMXP) to be discovered \citep{Wijnands1998}. Lying some 3.5~kpc distant \citep{Galloway2006}, the neutron star spins with a frequency of $\sim$401~Hz and is thought to be orbiting a low-mass brown-dwarf companion \citep{Bildsten2001} with a 2.01~hr orbital period \citep{Chakrabarty1998}. The established picture is that such AMXPs are recycled to millisecond periods from more slowly rotating neutron stars, due to the transfer of angular momentum from the accretion flow provided by the companion \citep{Bhattacharya1991}. In this sense, the AMXPs bridge the gap between low mass X-ray binaries (LMXB) and rotation powered millisecond pulsars. 

The accretion flow in AMXPs differs from that in black hole X-ray binaries (BHXRB) due to the stellar magnetic field and the solid neutron star surface. The magnetic field of SAX J1808.4-3658 is $\sim10^8$ G \citep{Hartman2008} and this is sufficient to channel accreting material approaching within a few neutron star radii along field lines towards the magnetic poles. The funnelled gas passes through an accretion shock before falling on to the magnetic poles. X-ray observations of AMXPs made with the {\it Rossi X-ray Timing Explorer} (RXTE) and {\it XMM-Newton} (XMM) have discovered distinct spectral components originating from these sources. The accretion disc and neutron star surface contribute to a soft component up to $\sim$10~keV, and Comptonisation in the shock region (most likely of seed photons from the impact hot-spot) produces a harder component extending to $\sim$100~keV \citep{Gierlinski2002}. Recent {\it XMM-Newton} observations of SAX J1808.4-3658 have shown two distinct blackbodies at soft energies plus a hard component, interpreted as an accretion disc, a neutron star hot-spot and an accretion shock \citep{Patruno09, Papitto2009}.

In an analogous manner to reflection in BHXRB \citep{Fabian1989}, it is thought that Comptonised photons reflect off the thin optically thick accretion disc producing a Compton reflection hump and fluorescent iron K$a$ line emission at $\sim$6.4~keV (see \citealt{Cackett2009a} for a review of broadened iron lines in neutron star LMXBs). The shape of the line provides information about the distance of the line-forming region, since reflection occurring close to the compact object produces a characteristic asymmetric profile due to Doppler-broadening and gravitational redshift. Observations of SAX J1808.4-3658 have revealed evidence suggestive of such a broadened iron line \citep{Cackett2009b,Papitto2009}. Iron K$a$ lines are also of particular importance because they place a constraint on the magnetospheric radius. It is, as yet, unclear how the magnetic field of a magnetised star truncates the disc, or what effect the star's rotation has on the disc inner edge. 

The X-ray emission from AMXPs is pulsed due to misalignment of the spin and magnetic poles. As the hot-spot and magnetically channelled accretion column rotate around the spin axis, the projected area and the magnitude of the Doppler shift changes, causing emission to be pulsed at the spin frequency. It is therefore not unreasonable in this scenario to imagine that changes in the line shape and/or energy with phase might also occur. For instance, if the accretion column illuminates significantly different parts of the disc with phase, reflection occurring from approaching and receding parts of the disc should display a change in line energy with phase. Exploring the variation of the line properties with phase, as well as the phase dependent correlation between the line equivalent width (EW) (or normalisation) and the reflection hump, might be a valuable window on the formation of this spectral feature. The iron K$a$ feature in SAX J1808.4-3658 has been known about for many years \citep{Gierlinski2002}, and we now know that this feature might be relativistically broadened \citep{Cackett2009b,Papitto2009}. Therefore, it is of interest whether this feature is also pulsed at some level, which might be expected if the illuminating continuum is also pulsed as seen from the inner disc (which might be expected in certain geometries, e.g. where a part of the disc is shielded by the neutron star itself).

Using the correct orbital ephemeris and referencing the photon arrival times to the Solar
System barycenter, it is possible to define an arbitrary phase-zero and `fold' all of the incoming photons in to phase-bins representative of a full rotation of the neutron star. By examining the energy spectra of photons from individual phase bins, one can perform phase-resolved-spectroscopy.   A phase-resolved analysis was performed of the 1998 outburst \citep{Gierlinski2002}, and the self-consistent continuum reflection and iron-line were found to be consistent with having a constant value as a function of phase.
In this work, we produce phase-resolved spectra of the best-sampled and best-studied 2002 outburst of SAX J1808.4-3658 as observed by {\it RXTE} over the `slow-decay' of the outburst \citep{Hartman2008} where pulse profiles are most stable and we can investigate the phase-resolved spectrum with high signal-to-noise. We shall explore changes in the iron-line properties with phase, and investigate the viability of the currently accepted picture of the broadened iron-line emission in AMXPs. 

\section{Observations and Data Reduction}
\label{obs}
We have concentrated on RXTE observations of SAX J1808.4-3658 from the slow-decay phase (MJD 52564-52574) of the 2002 outburst where the pulse profile is the most stable and the data is best sampled (in date order OBS ID 70080-01-01-02 to 70080-01-03-03 inclusive). The reader is referred to Figure 3 of \citet{Hartman2008} for a light curve of the outburst illustrating the data segment we used. For simplicity, only data taken with the { \sc E\_125US\_64M\_0\_1s} event mode of the {\it RXTE} Proportional Counter Array (PCA) were used in this work. Although during these observations different PCA detectors were switched on, we have only analysed data from PCU 2 which was always on, in order to ensure consistency. Gain and offset corrections, which attempt to match the energy scales in different PCU detectors, assign zero response in different channels of different PCUs \citep{Jahoda1996}. Combining different detectors involves weighting different detector responses, and this introduces unmodelled scatter in to the data. In total, 64 event files were analysed with a total exposure of 167~ksec, which provided sufficient signal to noise for phase-resolved spectroscopy using 16 phase bins. 

The photon arrival times in each event file were corrected to the Solar system barycenter with the ftool {\sc faxbary}. The photon phases were determined by using a high-order polynomial to ``whiten'' the phase residuals and increase the signal to noise of the pulsations. Additionally, the arrival time of photons was corrected for the orbital motion of the binary using a first-order approximation as outlined, for example, in \citet{Papitto2005}. A folded pulse profile was created for each separate event file, and fitted with a simple constant offset plus a sinusoid. During the slow-decay, pulse profiles have been shown to be stable and nearly sinusoidal \citep{Ibragimov2009, Hartman2008}. By fitting each event file with the best fitting sinusoidal function, an offset could be determined for each event file to accurately align the different files in phase for subsequent summation. Custom code was written to select the photons from PCU2 only and produce separate FITS files containing events from each of the 16 different phase bins for each event file. The ftool {\sc seextrct} was then used to produce PHA spectra for each phase bin.    

Background files were created using the ftool {\sc pcabackest}, and extracted using {\sc saextrct} over contemporaneous good time intervals created using the ftool {\sc maketime}. The ftool {\sc mathpha} was used to sum pha files from each respective phase bin and sum the background spectra. A weighted PCU2 response file was created (due to any variations in the response over time) from all observations using {\sc pcarsp} and {\sc addrmf}. Care was taken to adjust the exposure of the phase-resolved spectra to reflect the fact that each spectrum only contained 1/16th of the total exposure. Phase averaged PCU2 spectra were also obtained from the same observations.  

The pre-processed High Energy X-ray Timing Experiment (HEXTE) cluster A spectra from the same observations were summed together to provide some constraints on the phase-averaged power-law spectral component at higher energies.  

Throughout this work, spectral fitting was performed using {\sc XSPEC} version 12.5.1 \citep{Arnaud1996} with a 0.5\% systematic error for PCA and HEXTE data during fitting. It should be noted that the PCU 2 data has zero response in channel 10, but these counts are redistributed to other channels, hence the gap in the data in Figure \ref{fig:ironlineratio}.

\section[]{Analysis and Results}

\subsection{Phase-averaged spectrum}
\label{phaveraged}
Phase-averaged spectra of SAX J1808.4-3658 from the slow-decay phase of the 2002 outburst have been fitted before \citep{Ibragimov2009}. It is therefore known that a simple power-law fit shows residuals that are consistent with Compton reflection and blackbody hotspot emission. In both the 1998 and 2002 outbursts, thermal Comptonisation models have typically shown photon indices of $\sim1.8$ with blackbody temperatures around 0.7~keV and hot-spot areas in the range 20-110 $\rm{km}^2$ \citep{Gierlinski2002,Ibragimov2009,Poutanen2003}. 

\begin{figure}
\includegraphics[width=84mm]{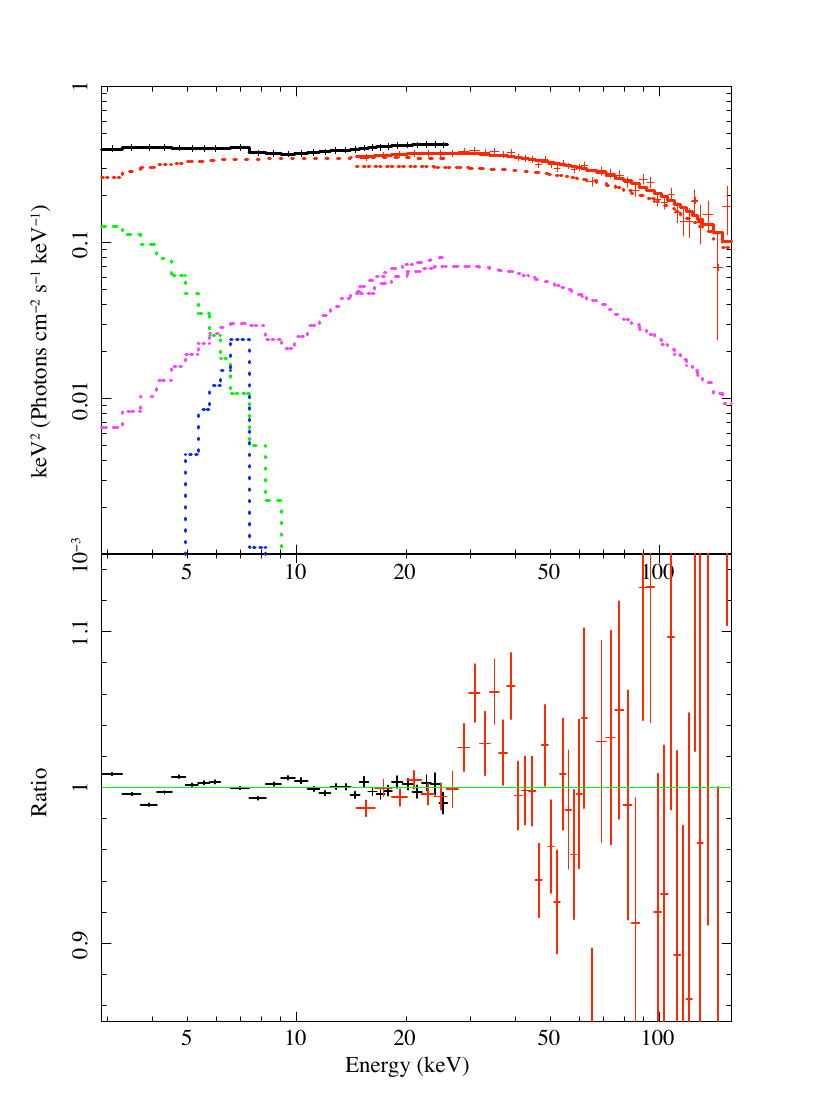}
\caption{Top: Unfolded phase-averaged spectrum (model multiplied by the ratio of observed data to the response folded spectral model) showing iron line, blackbody and reflected power-law components using the {\sc cons$\times$phabs$\times$(nthcomp+bbodyrad+rdblur$\times$(pexriv+gaussian))} model. Bottom: Data to model ratio plot of this fit.}
\label{fig:eeufspecphav}
\end{figure}

As well as fitting PCA data over 3.0 to 25.0~keV, the HEXTE spectra are also fitted simultaneously over the range 15.0-150~keV to constrain the power-law continuum and cut-off energy. The continuum was best-modelled using the thermal Comptonisation model {\sc nthcomp} \citep{Zycki1999}, and reflection was modelled using {\sc pexriv} \citep{Magdziarz1995} and {\sc gaussian} components convolved with {\sc rdblur}. Only the reflection component of the {\sc pexriv} model was included. A thermal Comptonisation model is chosen since the hot electrons in the accretion shock are thought to Compton up-scatter the seed photons from the hot-spot, whilst the blurred Gaussian is chosen to model the iron-line feature. All parameters were tied between the two instruments apart from a normalising constant to allow for a calibration offset. The hotspot emission is modelled using a {\sc bbodyrad} component. The {\sc bbodyrad} model is a single temperature blackbody spectrum parameterised by the temperature and the normalisation $\rm{R_{km}^{2}/D_{10}^{2}}$, where $\rm{R_{km}}$ is the source radius in km and $\rm{D_{10}}$ is the distance to the source in units of 10~kpc. The {\sc nthcomp} model is parameterised by the asymptotic power-law photon index, the electron temperature, the seed photon temperature and the normalisation (the photon flux defined at 1~keV in units of photons~cm$^{-2}$s$^{-1}$keV$^{-1}$). The {\sc pexriv} model describes an exponentially cut-off power-law spectrum reflected from ionised material and is parameterised by the power-law photon index, the power-law cut-off energy, the disc temperature, the disc ionization parameter \citep{Done1992}, the normalisation (photon flux at 1 keV) of the power-law only (without reflection) and a dimensionless reflection scaling factor. {\sc rdblur} is a convolution model taking account of the blurring due to relativistic effects from an accretion disc around a non-rotating black hole, and is parameterised by the power-law index of emissivity, the disc inner radius, the disc outer radius and the inclination. Full parameters of the fit are shown in Table \ref{tab:phaseavfit} and the unfolded spectrum is included in Figure \ref{fig:eeufspecphav}. The spectral fits reflect the average parameters over the 167~ksec of data we analysed from the slow-decay phase of the outburst. The parameters of this `average' fit are consistent with previous work, though it should be noted that as the mass accretion rate drops throughout the outburst, the disc inner edge likely recedes \citep{Ibragimov2009}, causing changes in the fit parameters. The fits were insensitive to $\xi$, the ionisation parameter, and the disc temperature so these were frozen at values of 100 erg~cm $\rm{s}^{-1}$ and $10^6$~K respectively. The low value of $\xi$ is consistent with the line energy which was constrained to lie between 6.4 and 6.9~keV and pegged at the lower limit during fitting. In order to avoid the {\sc pexriv} power-law cut-off energy becoming unphysical, this parameter was constrained to be equal to three times the value of the electron temperature. Freeing the seed photon temperature for Comptonisation from the blackbody temperature had no statistically significant effect, consistent with the notion that the seed photons for Comptonisation come from the hotspot emission. It should also be noted that the {\sc pexriv} normalisation and reflection fraction are degenerate parameters in these fits hence the large absolute values of reflection fraction quoted in Table \ref{tab:phaseavfit}. The relatively poor chi-squared value of the model quoted in Table \ref{tab:phaseavfit} is due to the higher energy HEXTE data. Removing the HEXTE data and fitting only PCA data results in a $\chi^2$/d.o.f of 16.56/18, but it was considered important to include the HEXTE data during fitting to better model the electron cut-off and thermal Comptonisation continuum. Grouping the HEXTE data in to bins of 4 channels each in this fit improves the $\chi^2$/dof to 38.44/28, a perfectly acceptable fit given the low number of degrees of freedom. Previous pulse-profile modeling \citep{Poutanen2003} has derived a lower limit of $65^\circ$ for the inclination of this source, but since no eclipses or absorption dips are seen, a value of $60^\circ$ is assumed throughout this work consistent with previous analysis \citep{Gierlinski2002, Ibragimov2009}. Any difference in inclination only introduced a systematic shift in the normalisation of spectral components, and therefore did not effect the relative phases of any pulsations.

\begin{table}
\centering
\caption{\label{tab:phaseavfit} Best fitting parameters for the phase-averaged {\sc cons$\times$phabs$\times$(nthcomp+bbodyrad+rdblur$\times$(pexriv+gaussian)} model used in section \ref{phaveraged}}.
\begin{tabular} { l | r}
\hline
\hline
Parameter & Value \\
\hline
$\rm{BB}_{kT}$(keV) & $0.676^{+0.02}_{-0.01}$ \\ [1ex]
$\rm{BB}_{norm}$ & $130.89^{+10.12}_{-12.87}$ \\ [1ex]
$\Gamma$ & $1.96^{+0.03}_{-0.03}$ \\ [1ex]
nthComp $\rm{KT}_{e}$ (keV) & $45.53^{+14.08}_{-8.10}$ \\ [1ex]
$\rm{nthComp}_{norm}$ & $0.051^{+0.004}_{-0.004}$ \\ [1ex]
foldE(keV) & $136.59^{+42.24}_{-24.30}$ \\ [1ex]
$\rm{PEXRIV}_{norm}$ & $0.034 \rm{(Frozen)}$ \\ [1ex]
$\rm{PEXRIV}_{refl}$ & $-3.96^{+0.93}_{-0.69}$ \\ [1ex]
$\rm{PEXRIV}_{\xi}$ & $100 \rm{(Frozen)} $ \\ [1ex]
$\rm{HEXTE}_{const}$ & $0.88^{+0.01}_{-0.01}$ \\ [1ex]
$\rm{R_{in}}$ $\rm{(R_G)}$ & $18.41^{+29.15}_{-12.41*}$ \\ [1ex]
Emissivity & $-3.30^{+1.1}_{-6.7*}$ \\ [1ex]
$\rm{Fe~line~E}$ (keV) & $6.40^{+0.08}_{-0.00*}$\\ [1ex]
$\rm{Fe~line~width}$ (keV) & $0.01 \rm{(Frozen)}$\\ [1ex]
$\rm{Fe~norm}$ $(10^{-4})$ & $6.11^{+2.70}_{-3.04}$ \\ [1ex]
$\chi^2$ / d.o.f & 97.18 / 62\\  [1ex]
\hline
\end{tabular}
\begin{flushleft}  Inclination was fixed at $60^{\circ}$, $R_{out}$ fixed at 1000 $R_G$, nH fixed at $0.113\times10^{22}$ and solar abundances assumed. Upper/lower error limits labelled with an asterisk show that the fit parameters were pegged, at times indicating that the fit was not particularly sensitive to this parameter.
\end{flushleft}
\end{table}

\subsection[]{Phase-resolved energy spectra}
\subsubsection{Evidence for a varying iron-line EW with phase}
\begin{figure}
\includegraphics[width=84mm]{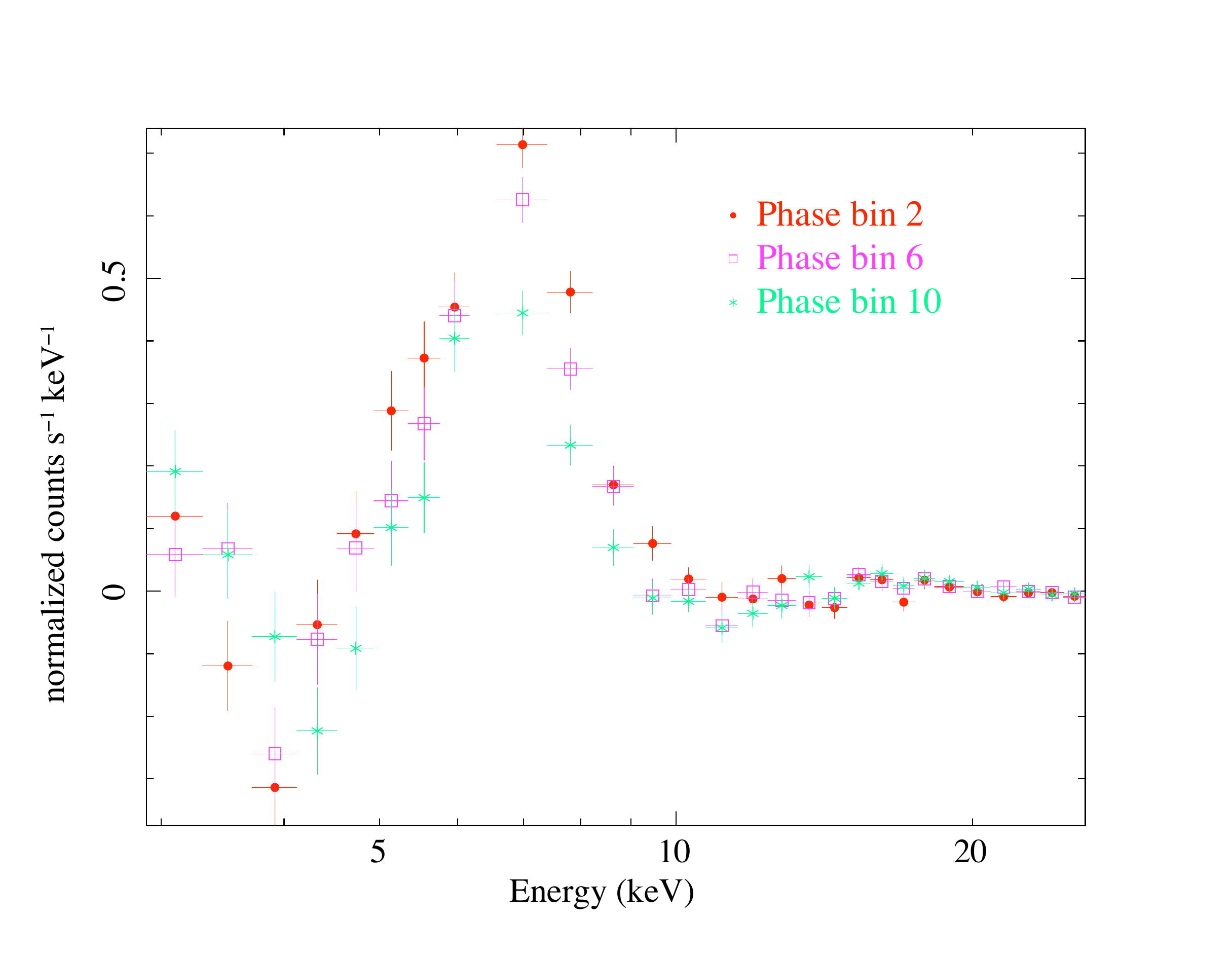}
\caption{Data to {\sc phabs*(pow+bbodyrad)} model residuals showing a broad excess in the iron line region. For clarity only three of the sixteen phase bins are shown here. The slight soft excess is most likely due to the Wien tail of accretion disc emission. The zero point from PCU channel 10 has been removed from this plot as discussed in the last paragraph of section \ref{obs}.}
\label{fig:ironlineratio}
\end{figure}

A simple phenomenological fit was performed to the 16-bin phase-resolved data consisting of a {\sc phabs*(pow+bbodyrad)} model. The absorption was fixed at $0.113\times10^{22}~{\rm cm}^{-2}$ as in \citet{Ibragimov2009} (obtained from the HEADAS tool {\it nh}) and the fit performed over the region 3.0-25.0~keV whilst ignoring the iron-line region of 5.5-10.0~keV. Component normalisations were allowed to be free for each of the 16 different spectra. Having fitted a model excluding the data in the iron-line region, by then including the data in this region one can plot the data-to-model residuals as in Figure \ref{fig:ironlineratio}. The plot of the residuals demonstrates a broad iron-line feature and hints at a changing iron-line equivalent width with phase. 

In the following sections, we shall explore different models to explain this apparent pulsation of the iron-line with phase. 

\subsubsection[]{Phenomenological Fits} 
\label{ironline}
\begin{figure}
\includegraphics[width=84mm]{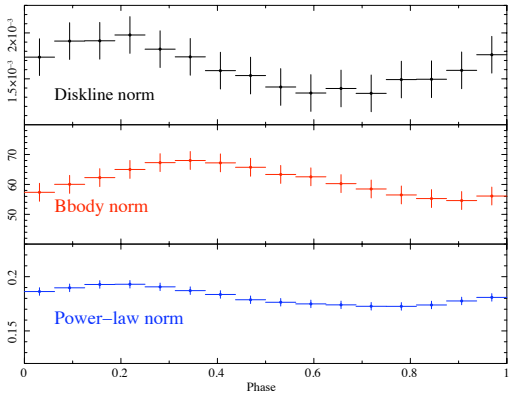}
\caption{Phenomenological fit demonstrating the pulsation in the {\sc diskline} normalisation with phase. The blackbody temperature was tied between phase bins in this fit. The approximate peak-to-peak change in the line normalisation was 42\%, in the blackbody normalisation 20\% and in the power-law normalisation 11\%.}
\label{fig:phenomlineuntied}
\end{figure}

\begin{figure}
\includegraphics[width=84mm]{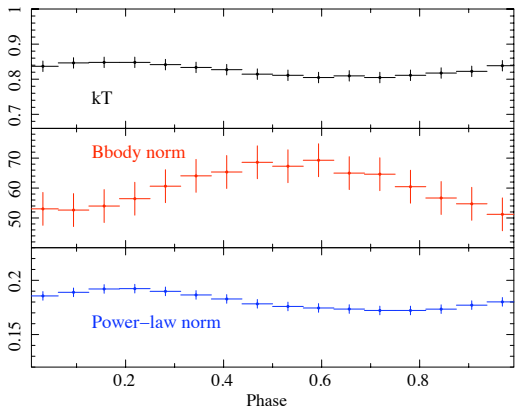}
\caption{Phenomenological fit demonstrating the pulsation in the blackbody temperature when {\sc diskline} normalisation is tied between phase-bins. The approximate peak-to-peak change in the blackbody temperature was 5\%, in the blackbody normalisation 28\% and in the power-law normalisation 11\%}
\label{fig:phenomktuntied}
\end{figure}

A fit was performed with the model {\sc phabs*(cutoffpl+bbodyrad+diskline)}, using the 16 phase-resolved spectra over 3.0 to 25.0~keV and the phase-averaged {\it HEXTE} spectrum over 15.0 to 150.0~keV. The iron line energy was constrained to lie within 6.4 and 6.9~keV and the inclination fixed at  $60^{\circ}$ as in \citet{Gierlinski2002} and \citet{Ibragimov2009}. With the cut-off power-law and blackbody normalisations free to vary between phase-bins (untied), and the {\sc diskline} normalisation constrained to its best fitting value across all phase bins (tied), the fit gave a $\chi^2$/dof of 756.52/467. The best fitting {\sc diskline} $\rm{R}_{in}$ was 11.9~$\rm{R}_g$ and the low value (1.63) of power-law photon index (compared to the phase-averaged fits) points to some degeneracy between the power-law and blackbody contributions at soft energies in this simple fit. Untying the {\sc diskline} normalisation between phase-bins improved the $\chi^2$/dof to 694.46/451 and the spectral components demonstrate the pulsations shown in Figure \ref{fig:phenomlineuntied}. Untying the {\sc diskline} normalisation simply allows the flux from the line to change with phase, rather than assuming the flux is constant as a function of phase. If the blackbody temperature is untied between phase-bins instead of the {\sc diskline} normalisation, the $\chi^2$/dof improves to 691.38/451 and the spectral components this time pulsate as shown in Figure \ref{fig:phenomktuntied}. Untying the blackbody temperature produces an $\sim80^{\circ}$ phase-shift in the pulsation of the blackbody normalisation compared to Figure \ref{fig:phenomlineuntied}, suggesting that there is some offset between the Doppler shift of blackbody photons (that affects the temperature measured by an observer at infinity and is dependent on the maximum line of sight velocity) and the projected area of the hot-spot which dominates the blackbody normalisation. It is also interesting to note that the {\sc diskline} normalisation and kT are in phase in Figures \ref{fig:phenomlineuntied} and \ref{fig:phenomktuntied} respectively. Again, it should be pointed out that the poor $\chi^2$/dof is due to the higher energy HEXTE data used to constrain the parameters in this fit. If a fit is performed excluding the HEXTE data (fixing the now unconstrained high energy cut off to the phase-averaged values) the $\chi^2$/dof improves to 473.58/407. Although the fits obtained with this simple phenomenological model are formally unacceptable, they do provide some physical insight in to the nature of the pulsations in this source which are discussed below. The detailed reflection models explored in section \ref{reflection} produce much more acceptable values of reduced $\chi^2$.

\begin{figure}
\includegraphics[width=84mm]{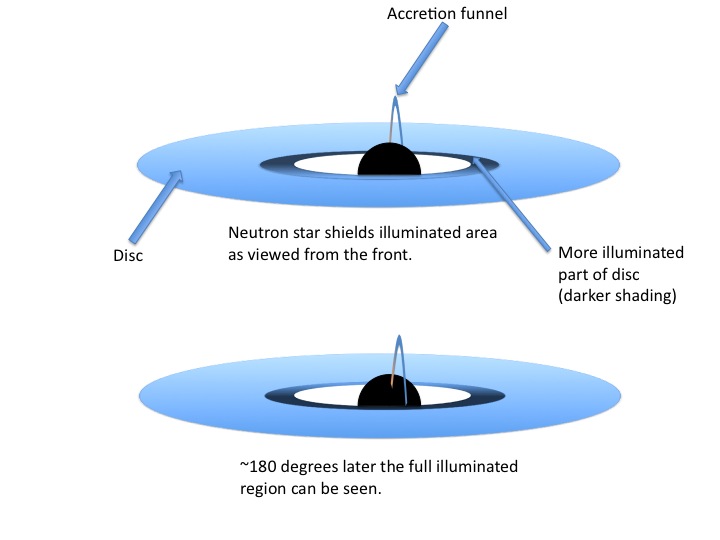}
\caption{Shielding of the reflected emission by the neutron star itself could lead to a variation in the visible area of line emission}
\label{fig:shielding}
\end{figure}

\begin{figure}
\includegraphics[width=84mm]{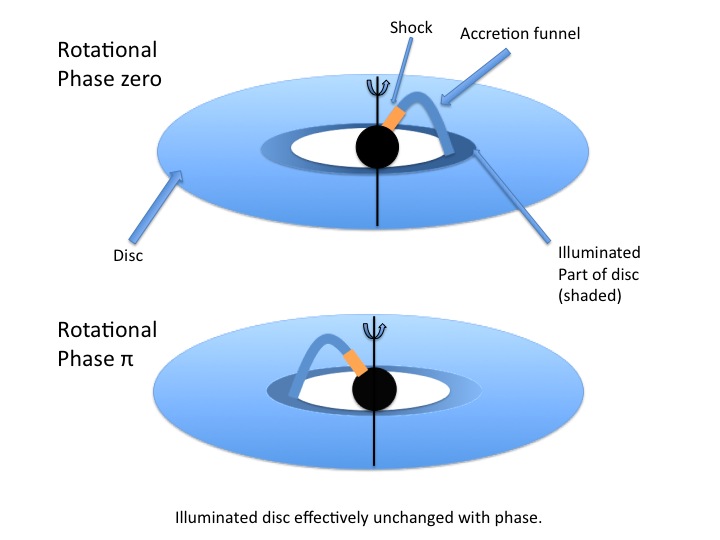}
\caption{Illustration of a geometry that could give rise to a constant reflection and iron line emission with phase. The small angle between the shock and the rotation axis means that the shock does not preferentially illuminate parts of the disc with different line of sight velocities. The illuminated area of the inner disc is illustrated using darker shading.}
\label{fig:discillum}
\end{figure}

This preliminary analysis leads to two explanations for the spectral pulsation, which are equally likely on statistical grounds alone. In the first case, the line normalisation is pulsating due to solid angle changes of the line emitting region and in the second case the line normalisation itself is constant, but the changing spectral shape of the blackbody component (governed by temperature changes, possibly due to Doppler effects) introduces a modulation at the spin frequency. It is important to point out here the physical situation that these alternative models are describing. In the first case, which invokes a varying line normalisation, the visible area of line emission changes with phase. This could, for instance, be due to the neutron star self-shielding the line emission from the disc as the system rotates, as shown in Figure \ref{fig:shielding}, or due to solid angle changes of the emitting area with phase. In the second scenario, where the line emission is constant with phase, the alignment of the spin and magnetic poles are such that the illumination of the disc by the accretion shock is more or less constant with phase as depicted in Figure \ref{fig:discillum}. The amount of observed iron-line emission is therefore fairly constant with phase, and the modulation can be explained solely in terms of the spectral changes in the blackbody component occurring in the same energy channels as the iron-line.

\subsubsection[]{Fits with reflection models}
\label{reflection}
\begin{figure}
\includegraphics[width=84mm]{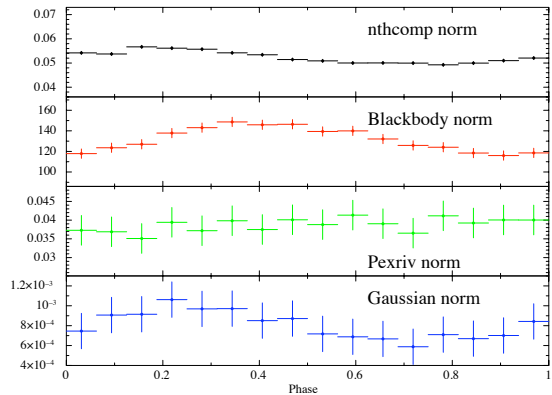}
\caption{Pulsations in the normalisations of spectral components using the physical reflection model described in the phase-averaged fits. Vertical axes are plotted to roughly show the relative amplitude of pulsations and the iron-line is modelled using a Gaussian component.}
\label{fig:fitoneparams}
\end{figure}

Having established these two alternative scenarios using the phenomenological fits, the same physically motivated reflection model described in the phase-averaged fits was fitted to the phase-resolved spectra (and {\it HEXTE} spectra). In the first fit, the normalisations of the {\sc bbodyrad}, {\sc nthcomp}, {\sc pexriv} and {\sc gaussian} components were untied between phase-bins, whereas the blackbody temperature and seed photon temperature were tied between phase-bins (and tied to each other). Throughout these phase-resolved fits, all parameters constrained to not vary with phase remained very close to the phase-averaged values quoted in Table \ref{tab:phaseavfit}. The $\chi^2$/dof for this fit was 455.0/435 and the pulsation of the spectral components with phase can be seen in Figure \ref{fig:fitoneparams}.

It is clear from Figure \ref{fig:fitoneparams} that, in this model, the gaussian (iron-line) normalisation is systematically varying with phase, whilst the reflection normalisation shows no clear evidence of doing so. We jointly fitted both the {\sc gaussian} and {\sc pexriv} normalisations shown in Figure \ref{fig:fitoneparams} with a simple constant offset plus a sinusoid, which forced the fractional amplitude and phase of the normalisation variations to be the same. The fit was poor, but was significantly improved by untying the phase ($\Delta\chi^{2}=3.23$ for one additional free parameter) and amplitudes ($\Delta\chi^{2}=4.69$ for one additional free parameter). The likelihood of an observed improvement by chance was 0.003 and $3.7\times10^{-6}$ for untying phase and then fractional amplitude respectively. The fractional amplitude of variation in the {\sc pexriv} normalisation was close to zero i.e. consistent with a constant {\sc pexriv} reflection component. This inconsistency in the reflection components is the most compelling evidence that the ``iron line pulsation'' is actually no more than a change in the underlying blackbody spectral shape with phase. The reflection and iron-line normalisation should be strongly correlated in amplitude and in phase with one another because they both emerge as a consequence of hard X-ray irradiation of the disc. In fact, Figure \ref{fig:fitoneparams} demonstrates no discernible pulsation in the {\sc pexriv} component at all. 

Untying the blackbody temperature between phase-bins in this fit did not significantly improve the fit. 

\begin{figure}
\includegraphics[width=84mm]{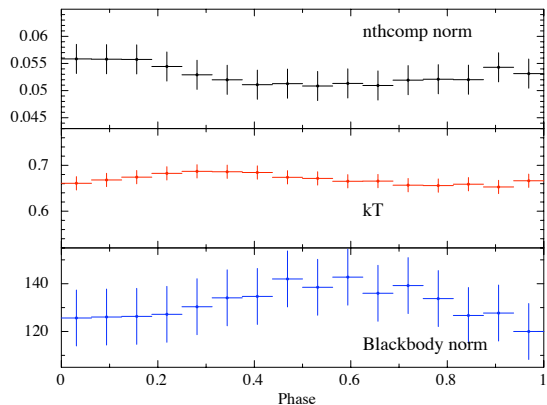}
\caption{Pulsations in the {\sc bbodyrad} and {\sc nthcomp} normalisations and the blackbody temperature with phase with all reflection parameters fixed between phase-bins.}
\label{fig:fittwoparams}
\end{figure}

\begin{figure}
\includegraphics[width=84mm]{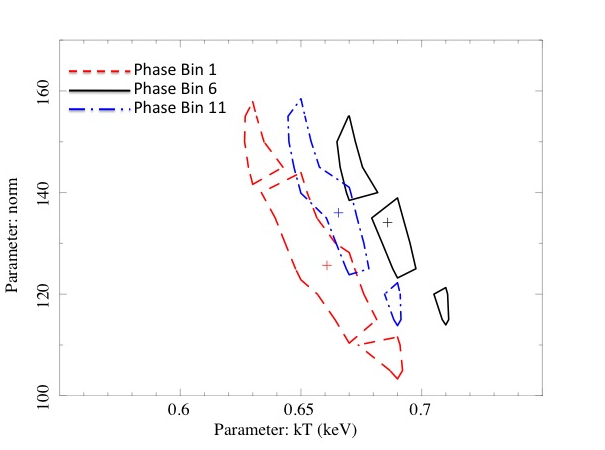}
\caption{99.9$\%$ confidence contours of blackbody normalisation versus kT corresponding to phase bins 1, 6 and 11 in figure \ref{fig:fittwoparams}. Due to the coarseness of the grid used to search parameter space (a step size of 0.02 keV and 5 for kT and normalisation respectively) the contours do not join up.}
\label{fig:contour}
\end{figure}

A second fit was performed to the phase-resolved spectra using the same model described in the phase-averaged fits, but this time constraining the reflection and line normalisations to be constant with phase. The $\chi^2$/dof of this fit was 492.46/467. Allowing the blackbody temperature to also vary with phase (it was still assumed that the blackbody temperature and seed photon temperature for Comptonisation were identical) improved the $\chi^2$/dof to 464.07/451 with an f-test probability of 0.039. The variations with phase are shown in Figure \ref{fig:fittwoparams}. Again, as with the earlier phenomenological fits, the blackbody normalisation is shifted along in phase when the blackbody/seed photon temperature is a free parameter. Figure \ref{fig:contour} shows 99.9$\%$ confidence contours for three phase bins (1,6 and 11) of blackbody temperature against blackbody normalisation and demonstrates that the temperature and normalisation are correlated. However, the fact that the 99.9$\%$ contours do not overlap significantly in temperature for a given normalisation (and vice versa) and there is no anti-correlation in the pulsations of these components in figure \ref{fig:fittwoparams}, suggests that the pulsation in these components might be real. 

Untying the seed photon temperature for thermal Comptonisation and the blackbody temperature had no significant effect on the spectral fits. 

There is no concrete statistical evidence to favour blackbody temperature variations over varying reflection, but there is a good physical reason to do so as outlined above. The best fit arises by invoking a changing blackbody temperature with phase and, although the error bars are large, the amplitude and shape of reflection variations do not match the line variations seen in Figure \ref{fig:fitoneparams}.  In the next section we will show that the observed variations can be explained quite simply in terms of the Doppler shifts (to produce the observed hotspot temperature variation) and solid angle changes (to produce the normalisation variation) which are expected for this system.

Table \ref{tab:rmsvalues} shows the fractional rms values for the different model components of each model described above.  

\begin{table}
\centering
\caption{\label{tab:rmsvalues} Table of fractional rms values for different model components}
\begin{tabular}{ | c | c | c | c | c |}
\hline
\hline
\multicolumn{5}{| c | }{Fractional RMS}\\
\hline
&Fig 3&Fig4&Fig7&Fig8\\[1ex]
\hline
Diskline & 0.157 & n/a & n/a & n/a \\ [1ex]
Bbody norm & 0.072 & 0.10 & 0.083 & 0.046 \\ [1ex]
Power-law & 0.039 & 0.039 & n/a & n/a \\ [1ex]
kT & n/a & 0.018 & n/a & 0.015 \\ [1ex]
nthcomp norm & n/a & n/a & 0.044 & 0.030 \\ [1ex]
Pexriv norm & n/a & n/a & 0.020 & n/a \\ [1ex]
Gauss norm & n/a & n/a & 0.155 & n/a \\ [1ex]
\hline
\end{tabular}
\begin{flushleft} 
Fractional RMS values are obtained from the best-fitting constant offset plus sinusoid model. Quoted value for the Pexriv normalisation is an upper limit.
\end{flushleft}
\end{table}

\section{Discussion}
\label{discussion}

\begin{figure}
\includegraphics[width=84mm]{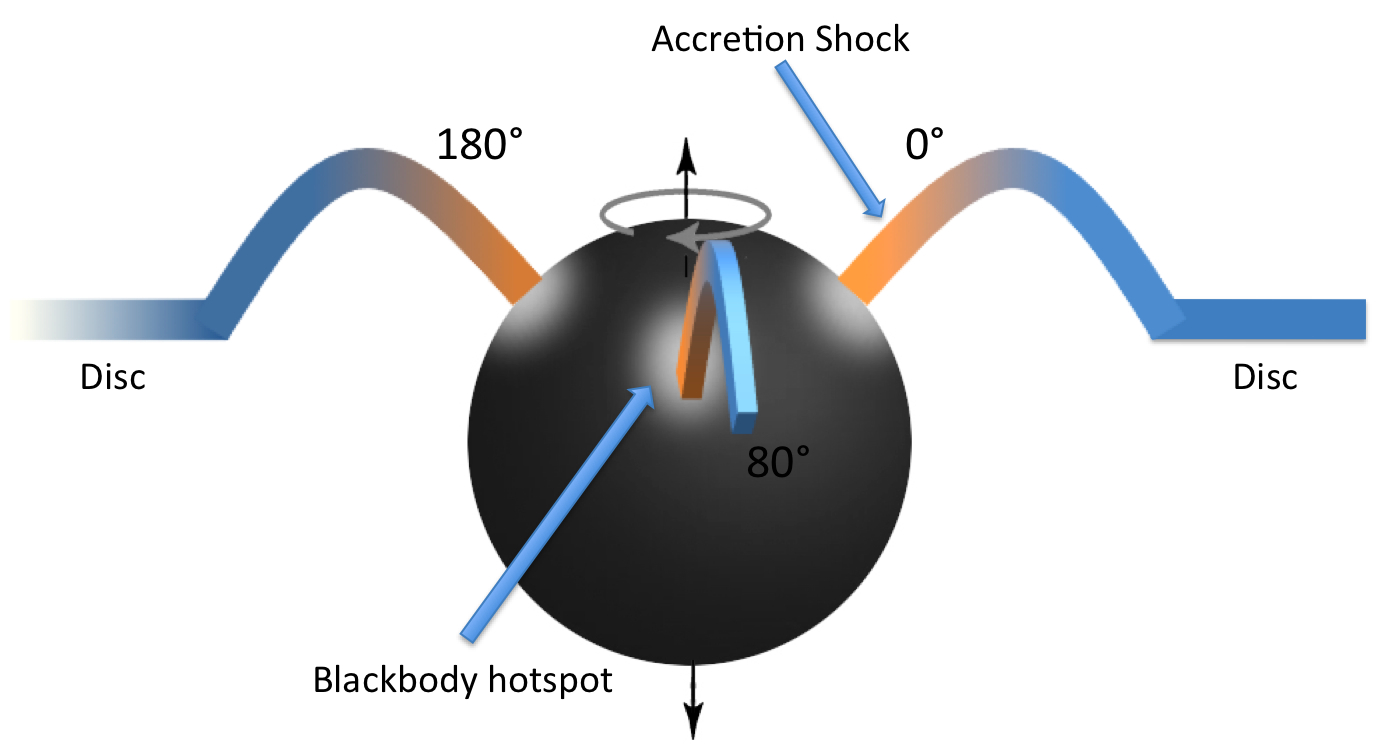}
\caption{The projected area of the blackbody hotspot changes more than the projected area of the accretion shock region as the neutron star rotates. A snapshot at three different rotational phases are included on this diagram.}
\label{fig:phase}
\end{figure}

The rms values shown in Table \ref{tab:rmsvalues} for the physically motivated reflection model invoking blackbody temperature variations (Figure \ref{fig:fittwoparams}), suggest a larger variation in the blackbody component than the Comptonised component. As shown in Figure \ref{fig:phase}, this is most likely due to blackbody emission arising from a flatter, more horizontally extended `pancake' type region compared to the Comptonised emission from the accretion shock which could extend vertically from the neutron star surface. As the star rotates, the projected area of the blackbody region changes more than the accretion shock region, leading to a greater amplitude of variation.  The fractional variation from the Comptonised emission might also be reduced if the accretion column is relatively optically thin (which mitigates the effect of solid angle variation), or if there exists a second, constant Comptonising region, e.g. an extended corona. The pulsation in the Comptonised component is significantly out of phase with the blackbody component, as seen in previous work \citep{Gierlinski2002,Ibragimov2009}. However, the phase shift of $\sim$$170^{\circ}$ between these components in Figure \ref{fig:fittwoparams} (when kT is a free parameter), is much larger than in previous work where the shift was $\sim$$50^{\circ}$ \citep{Gierlinski2002} and $\sim$$70^{\circ}$ \citep{Ibragimov2009} with kT fixed. By fixing the blackbody temperature, the offset between the blackbody and Comptonised component is found to be $\sim50^{\circ}$, consistent with \citet{Gierlinski2002}. It would therefore appear that some of the role of changing normalisation is subsumed by the changing kT, shifting the variations in blackbody normalisation to later phases. It is perhaps surprising how sensitive the phase-shift of the blackbody component pulsations are to the choice of model used. If the accretion column is optically thick and vertically elongated above the neutron star surface (as argued above), then the solid angle of the column could be maximised when it points away from the observer. This would coincide with the minimum projected hotspot area and can therefore explain the $\sim$$180^{\circ}$ phase shift between the blackbody and Comptonised components. 

For a 15~km neutron star, the Doppler shift $\Delta$$E$$/E$$=2\pi$$R\nu/c$, produces a large shift at the equator (e.g., around 13\% at a rotation frequency of 401Hz). For an accretion shock aligned at $10^\circ$ to the rotation axis, the upper limit obtained by \citet{Ibragimov2009}, R is reduced from 15~km at the equator to just 2.6~km and the Doppler shift reduced to just 2\%. If we discount the possibility of pulsed iron line variations (as implied by the much weaker variations in the reflection continuum), the phase-resolved fits suggest that some of the pulsed variation in the spectral shape can be explained by a varying Doppler shift of the blackbody photons underlying the iron-line. If we associate the blackbody temperature change with a Doppler shift, Figure \ref{fig:fittwoparams} demonstrates a peak-to-peak change in temperature of approximately 0.035~keV which in this source, assuming an inclination of $60^\circ$, translates to a radius of rotation of 3.5~km, consistent with the arguments of \citet{Poutanen2003}. This shift is consistent with a hot-spot aligned at approximately 13.5$^{\circ}$ (20$^{\circ}$) to the rotation axis, for a 15~km (10~km) radius neutron star. The fact that the shock makes a small angle with the rotation axis of the neutron star implies that the whole inner disc is illuminated more or less evenly i.e. there is no preferential illumination of different parts of the disc with different line of sight velocities which would lead to a change in the iron-line shape with phase.

A phase shift measured as $\sim80^{\circ}$ can be seen in Figure \ref{fig:fittwoparams} between the pulsation in the blackbody temperature (driven by Doppler variations) and the blackbody normalisation (driven by projected area effects). As dicussed in \citet{Gierlinski2002}, this is consistent with a geometry where Doppler variations are strongest when the hot-spot is at the limb of the star, and projected area effects are strongest $\sim90^{\circ}$ later when the solid angle of the hot-spot is maximised.  Clearly, this picture is only approximate, since we do not take account of effects such as light-bending (e.g. \citealt{Ibragimov2009}).

Our fits to these data are consistent with the idea that the iron-line shape and normalisation remain constant with phase, suggesting a geometry where the illumination of the disc by the accretion shock is fairly constant. By assuming that the {\it only} variable components are the normalisations of the blackbody and Comptonised components \citep{Gierlinski2002,Ibragimov2009}, one can detect an apparent pulsation in the residual iron-line component. However, any apparent change in the iron-line normalisation with phase can be explained by taking into account the Doppler boosting of the underlying blackbody continuum.  The constancy of the iron line is not really surprising, because for the small offset of the accretion column from the rotation axis that is envisaged here, the disc will not see much variation in solid angle of the accretion-shock region, and variations in Doppler beaming of the accretion-shock emission toward the disc will be small.  The presence of a constant power-law continuum component from a corona would also serve to reduce the variability of the iron line emission. 

\section{Conclusions}

Phase-resolved spectroscopy of the 2002 outburst of the AMXP SAX J1808.4-3658 as observed by {\it RXTE} reveals no convincing evidence for pulsations in the iron-line or changes of any line properties with phase. By allowing the blackbody spectral shape to change (i.e. freeing the blackbody temperature during fitting) one can allow for variable Doppler boosting of the underlying blackbody continuum, which introduces a larger phase-shift between the Comptonised component normalisation and the blackbody normalisation than has been seen in previous work. The phase-shift between the blackbody normalisation and the blackbody temperature is approximately consistent with the expected $90^{\circ}$ offset between the maximum projected area and maximum Doppler shifts. The fraction of the disc illuminated by the Comptonising region appears to be roughly constant with rotational phase. 

\section*{Acknowledgments}

TW is supported by an STFC postgraduate studentship grant, and PU is supported by an STFC Advanced Fellowship. The research leading to these results has received funding from the European Community's Seventh Framework Programme (FP7/2007-2013) under grant agreement number ITN~215215  ``Black Hole Universe". This research has made use of data obtained from the High Energy Astrophysics Science Archive (HEASARC), provided by NASA's Goddard Space Flight Center, and also made use of NASA's Astrophysics Data System. The authors would like to thank the anonymous referee for useful comments.

\label{lastpage}

\end{document}